\documentclass[%
 aip,
 amsmath,amssymb,
 reprint,%
]{revtex4-1}

\usepackage{graphicx}
\usepackage{dcolumn}
\usepackage{bm}
\usepackage[mathlines]{lineno}
\usepackage{amsthm}
\usepackage{amssymb}
\usepackage[table,xcdraw]{xcolor}
\usepackage[utf8]{inputenc}
\usepackage[T1]{fontenc}
\usepackage{mathptmx}
\usepackage{mathrsfs}
\usepackage{mathtools} 
\usepackage{etoolbox}
\usepackage{booktabs}
\usepackage[table,xcdraw]{xcolor}
\usepackage{tabularx}
\usepackage{float}
\usepackage{siunitx}
\usepackage{booktabs}   
\usepackage{multirow}   
\usepackage[referable]{threeparttablex}

\makeatletter
\def\@email#1#2{%
 \endgroup
 \patchcmd{\titleblock@produce}
  {\frontmatter@RRAPformat}
  {\frontmatter@RRAPformat{\produce@RRAP{*#1\href{mailto:#2}{#2}}}\frontmatter@RRAPformat}
  {}{}
}%
\makeatother
\newcolumntype{x}{>{$}l<{$}}  
\newcolumntype{y}{>{$}c<{$}}  
\newcolumntype{z}{>{$}r<{$}}  

\newcommand{\txtd}{\text{d}}
\newcommand{\dr}{\text{d}\ensuremath{\mathbf{r}}}

\newcommand{\occ}{\text{occ}}

\newcommand{\vext}{\ensuremath{v_\text{ext}}}

\newcommand{\Exc}{E_\text{xc}}
\newcommand{\xc}{\text{xc}}
\newcommand{\xcx}{\text{x}}
\newcommand{\xcc}{\text{c}}
\newcommand{\ext}{\text{ext}}
\newcommand{\boldr}{\ensuremath{\mathbf{r}}}

\newcommand{\Vee}
{\ensuremath{\hat{V}_\text{ee}}}

\newcommand{\ndep}{\ensuremath{[n]}}

\begin{document}
\preprint{AIP/123-QED}

\title{Coupling-Constant Averaged Exchange-Correlation Hole for He, Li, Be, N, Ne Atoms from CCSD}
\author{Lin Hou}
\affiliation{
Physics and Engineer Physics Department, Tulane University.
}
 
\author{Tom J. P. Irons}
\affiliation{
School of Chemistry, University of Nottingham, University Park, Nottingham NG7 2RD, United Kingdom.
}

\author{Yanyong Wang}
\affiliation{
Physics and Engineer Physics Department, Tulane University.
}

\author{James W. Furness}
\affiliation{
Physics and Engineer Physics Department, Tulane University.
}

\author{Andrew M. Wibowo-Teale}
\affiliation{
School of Chemistry, University of Nottingham, University Park, Nottingham NG7 2RD, United Kingdom.
}
\affiliation{
Hylleraas Centre for Quantum Molecular Sciences, Department of Chemistry, University of Oslo, P.O. Box 1033, N-0315 Oslo, Norway.
}

\author{Jianwei Sun}
\affiliation{
Physics and Engineer Physics Department, Tulane University.
}

\begin{abstract}
Accurate approximation of the exchange-correlation (XC) energy in density functional theory (DFT) calculations is essential for reliably modelling electronic systems. Many such approximations are developed from models of the XC hole; accurate reference XC holes for real electronic systems are crucial for evaluating the accuracy of these models however the availability of reliable reference data is limited to a few systems. In this study, we employ the Lieb optimization with a coupled cluster singles and doubles (CCSD) reference to construct accurate coupling-constant averaged XC holes, resolved into individual exchange and correlation components, for five spherically symmetric atoms: He, Li, Be, N, and Ne. Alongside providing a new set of reference data for the construction and evaluation of model XC holes, we compare our data against the exchange and correlation hole models of the established LDA and PBE density functional approximations. Our analysis confirms the established rationalization for the limitations of LDA and the improvement observed with PBE in terms of the hole depth and its long-range decay, demonstrated in real-space for the series of spherically-symmetric atoms.  

\end{abstract}






\maketitle 

\section{Introduction}\label{sec:intro}

Density functional theory (DFT) has become the most widely used electronic structure method on account of the high accuracy that can be achieved with relatively low computational scaling in the study of many-body systems. DFT is now the mainstay of electronic structure modelling across a range of areas including condensed matter physics, quantum chemistry and materials science. Through determining the ground-state electron density directly, the construction of a many-body wavefunction can be avoided, whilst allowing the ground-state energy and important chemical and physical properties of the system to be evaluated.\cite{Kohn1964/PhysRev.136.B864}

Most practical implementations follow the Kohn-Sham formulation of DFT (KS DFT), in which the density is represented by a set of one-electron orbitals that form a single determinant wavefunction.\cite{KohnSham1965} Besides being computationally convenient, this approach enables the largest contributions to the energy to be evaluated exactly, requiring approximation of the remaining exchange-correlation (XC) component, responsible for many-electron effects. It is therefore the quality of the XC density functional approximation (DFA) that determines the accuracy of the energy and other ground-state properties predicted by KS DFT.\cite{ParrYang1989DT}

One approach to the development of XC DFAs has been by modelling coupling-constant averaged XC hole, denoted for an electron of spin $\sigma_{1}$ at position $\mathbf{r}$ by $\bar{n}_{\mathrm{xc}}^{\sigma_{1}\sigma_{2}}(\mathbf{r},\mathbf{r+u})$, from which the XC energy may be evaluated as\cite{constantin2013construction}
\begin{equation}\label{eq:Exc_xch_express}
    E_{\text{xc}}\ndep = \sum_{\sigma_{1}\sigma_{2}} \iint\text{d}\boldr \text{d}\mathbf{u} \frac{n^{\sigma_{1}}(\boldr) \bar{n}_{\mathrm{xc}}^{\sigma_{1}\sigma_{2}}(\mathbf{r},\mathbf{r+u})}{ 2\vert\mathbf{u}\vert},
\end{equation}
in which $n^{\sigma_{1}}$ represents the $\sigma_{1}$-spin density. This may be resolved into separate exchange and correlation holes, denoted $n_x^{\sigma_{1}}$ and $\bar{n}_c^{\sigma_{1}\sigma_{2}}$ respectively , as\cite{constantin2013construction}
\begin{equation}\label{eq:xc_hole_resolved}
    \bar{n}_{\mathrm{xc}}^{\sigma_{1}\sigma_{2}}(\mathbf{r}, \mathbf{r}+\mathbf{u}) = n_x^{\sigma_{1}}(\mathbf{r},\mathbf{r}+\mathbf{u})\delta_{\sigma_{1}\sigma_{2}} + \bar{n}_c^{\sigma_{1}\sigma_{2}}(\mathbf{r},\mathbf{r}+\mathbf{u}).
\end{equation}
The accurate calculation of the XC energy therefore relies on the quality of the XC hole model, the assessment of which is instructive in understanding the limitations of XC DFAs when applied to real systems and developing more reliable models.\cite{Perdew2003DFTprimer} However, the tendency to directly model the XC energy density widely-used DFAs often overlooks the significance of XC holes, with the result that studies focusing on XC holes have been relatively limited. Notably, earlier successful DFAs, such as the PW91 functional by Perdew and Wang,\cite{PW91} and the construction of the more recent SCAN functional,\cite{Sun2015/scan} were based on modelling or understanding of the XC hole. Furthermore, the significance of XC holes is being increasingly recognized again in new DFA developments.\cite{Burke2020/CPDFT}

The benchmarking of model XC holes against accurate reference data remains a challenging yet essential endeavor in the development of DFAs. Numerous studies have explored the exact exchange hole for a small range of systems, including the jellium surface\cite{constantin2009many} and the H, He, Li, N and Ne atoms.\cite{constantin2006meta,constantin2013construction,ernzerhof1998generalized} Additionally, investigations have been conducted on the exact correlation hole at full interaction strength (as opposed to the coupling-constant averaged form most appropriate to model in DFT) for Hooke's atom and the He, Li, $\text{Be}^{2+}$, Be and Ne atoms.\cite{o2003wave,cioslowski1998electron} 

Evaluating accurate coupling-constant averaged XC holes is not a trivial task; the adiabatic connection (AC) must be constructed between the non-interacting and the physically-interacting limits, with the density remaining constant as the physically-interacting density throughout. This requires the effective potential to be optimized at each interaction strength such that it yields the physical density, for which the Lieb optimization provides a robust framework.\cite{Lieb1983,WYang1983/Lie_opt} In combination with a coupled cluster singles and doubles (CCSD) reference, the Lieb optimization has been used to study two-electron systems, particularly focusing on the XC energy of the Helium isoelectronic series.\cite{Teale2009,Teale2010} More recently, this approach has been employed by the present authors to investigate the CCSD cusp effect-driven error using the Hooke's atom as a model.\cite{hou2024capturing} These investigations provide valuable insights into the characteristics of XC holes and the effect of basis-set size on the cusp-driven errors in CCSD.

In the present work, we employ a similar methodology to study the XC holes in a set of five spherically symmetric atoms: He, Li, Be, N and Ne. Atoms in this set have between two and 10 electrons and this set contains both atoms with closed-shell and atoms with open-shell ground-state electronic configurations, allowing the relationship between number of electrons, electronic spin state and XC hole characteristics to be investigated.  Moreover, these atoms have been extensively studied in prior research \cite{o2003wave,cioslowski1998electron}, providing a valuable benchmark for the comparison and validation of the present work. By focusing on these five atoms, our objective is to develop a comprehensive understanding of XC holes and their distinctive features in a diverse range of atomic systems.


We commence in Section~\ref{sec:theory_method} by providing an overview of the theoretical framework for computing the coupling constant-dependent system and spherically-averaged XC holes for open-shell systems. The details of the calculations undertaken in this work are then discussed in Section~\ref{sec:computation}. In Section~\ref{sec:results}, we present the coupling-constant averaged XC holes and separate exchange and correlation holes for these five atoms evaluated using the Lieb optimization with a CCSD reference, examining the convergence of the energy with both basis set size and inter-electronic separation $u$ over which the holes are integrated. In addition, we study the convergence of the integrals of the exchange and correlation holes with respect to $u$ towards the known values of their respective integrals over all space: the \textit{sum rule} criteria for exchange and correlation holes. We apply this analysis to the model exchange and correlation holes of the local density approximation (LDA) and the Perdew-Burke-Ernzerhof (PBE) DFAs, comparing their characteristics with those of the accurate Lieb+CCSD data and rationalize their differences. We conclude our work with a brief summary in Section~\ref{sec:conclusion}.

\section{Theory and methodology}\label{sec:theory_method}
In KS-DFT, the ground-state energy of a many-electron system in an external potential $v_\ext(\boldr)$ is obtained by mapping the interacting system of electrons to an auxiliary non-interacting system of electrons with the same density. The problem is then reduced to a set of one-electron eigenfunction equations, the solutions to which are a set of orbitals in which the density is represented.~\cite{KohnSham1965} The ground-state energy as a functional of the electron density $n(\boldr)$ can be resolved into the sum of several contributions as
\begin{equation}\label{eq:KS-total}
    E\ndep = T_\text{s}\ndep + E_\text{H}\ndep + \Exc\ndep + \int\dr\, \vext(\boldr) n(\boldr),
\end{equation}
where $T_\text{s}$ is the non-interacting kinetic energy, which is evaluated exactly using the KS orbitals, and $E_\text{H}$ the classical electrostatic Hartree energy, which is evaluated exactly in terms of $n(\boldr)$. The only term in Eq.~(\ref{eq:KS-total}) which must be approximated is the XC energy $E_\xc\ndep$, which describes all of the many-electron effects in the system.

The KS non-interacting system may be linked to the physically-interacting system by continuously varying the strength of the electron-electron interaction between the non-interacting and physically-interacting limits, scaling the two-electron operator $\Vee$ by a coupling-constant $\lambda$ between zero and one. The electronic state evolves through a family of solutions to the $\lambda$-interacting Hamiltonian,
\begin{equation}\label{eq:H_lambda}
    \hat{H}_\lambda = \hat{T} + \lambda\Vee + \sum_i v_\lambda(\boldr_i),
\end{equation}
where $\hat{T}$ is the kinetic energy operator and $v_\lambda$ a modified external potential, thus establishing an adiabatic connection between the non-interacting and physically-interacting systems.~\cite{LangrethPerdew/adiabatic_connection} The modified external potential $v_\lambda$ is determined for each interaction strength such that the density remains constant at the physical ($\lambda=1$) density for all $\lambda$; it reduces to the local KS potential $v_\text{s}$ at $\lambda=0$ and is equal to the physical external potential $\vext$ at $\lambda=1$. 

Given the normalized ground-state $N$-electron wave function of the $\lambda$-interacting system $\Psi_\lambda$, the diagonal of the spin-resolved two-particle density matrix is expressed as~\cite{Davidson1976,McWeeny1960/advances_DMtheory}
\begin{align}
    n^{\sigma_1 \sigma_2}_{2,\lambda}\left(\boldr,\boldr'\right) &\equiv N(N-1)\sum_{\sigma_3,\cdots,\sigma_N} \int\dr_3 \cdots\int \dr_N \notag \\
    &\qquad\quad \left|\Psi_\lambda\left(\boldr\sigma_1,\, \boldr'\sigma_2,\, \boldr_3\sigma_3,\, \cdots, \boldr_N\sigma_N \right)\right|^2 .\label{eq:2nd_order_DM}
\end{align}
Whilst the two-particle density cannot be diagonalized by a unitary transformation of the one-electron orbitals,~\cite{Davidson1976} it may be used to evaluate the expectation value of two-body operators.~\cite{Perdew2003DFTprimer} 

The spin-resolved XC hole $n_{\text{xc},\lambda}^{\sigma_1 \sigma_2}\left(\mathbf{r}, \mathbf{r}'\right)$ is computed from the spin-resolved two-particle density matrix $n^{\sigma_1 \sigma_2}_{2,\lambda}$ and the spin densities $n^\sigma$ as
\begin{equation}
    n_{\text{xc},\lambda}^{\sigma_1 \sigma_2}\left(\mathbf{r}, \mathbf{r}'\right)=\frac{n^{\sigma_1 \sigma_2}_{2,\lambda}\left(\mathbf{r}, \mathbf{r}'\right)}{n^{\sigma_1}\left(\mathbf{r}\right)}-n^{\sigma_2}\left(\mathbf{r}'\right).
\end{equation}
At $\lambda=0$, the opposite spin component of XC hole $n_{\xc,\lambda=0}^{\sigma\sigma'}(\boldr, \boldr')=0$ while the same spin component reduces to the KS exchange hole, 
\begin{align}
    n^{\sigma\sigma}_\xcx(\boldr, \boldr') &= n^{\sigma\sigma}_{\text{xc},\lambda=0}(\boldr, \boldr') \notag \\
    &= -\frac{\sum^\occ_{i,\,j} \psi^*_{i\sigma}(\boldr) \psi_{j\sigma}(\boldr) \psi^*_{j\sigma}(\boldr') \psi_{i\sigma}( \boldr') }{ n^{\sigma}(\boldr)}, \label{eq:Xh_def}
\end{align}
where $\psi_{i\sigma}(\boldr)$ are the KS spin-orbitals. Therefore the spin-resolved $\lambda$-averaged correlation hole can be defined by
\begin{equation}
\begin{aligned}
    \bar{n}^{\sigma\sigma}_\xc( \boldr, \boldr')&=n^{\sigma\sigma}_\xcx\left(\boldr, \boldr'\right)+\bar{n}^{\sigma\sigma}_{\mathrm{c}}(\boldr, \boldr'), \\
    \bar{n}^{\sigma\sigma'}_\xc( \boldr, \boldr')&=\bar{n}^{\sigma\sigma'}_{\mathrm{c}}(\boldr, \boldr').
    \label{eq:Ch_def}
\end{aligned}
\end{equation}
The XC holes may be spherically-averaged by integration over the volume element of the second electron coordinate as
\begin{equation}
    n_{\text{xc}, \lambda}^{\sigma_1 \sigma_2}(\mathbf{r}, u)=  \int \frac{\mathrm{d} \Omega_{\mathbf{u}}}{4 \pi} n_{\text{xc}, \lambda}^{\sigma_1 \sigma_2}(\mathbf{r}, \mathbf{r}+\mathbf{u}),
\end{equation}
where $\mathbf{u}=\mathbf{r}'-\mathbf{r}$. The system-averaged spin-resolved XC hole is then evaluated by spatially integrating the spherically-averaged XC hole with the spin density of the reference electron, then dividing by the number of spin $\sigma_1$ electrons $N_{\sigma_{1}}$ as
\begin{equation}\label{eq:spin_nxc}
    \langle n_{\text{xc}, \lambda}^{\sigma_1 \sigma_2}\rangle(u)  = \frac{1}{N_{\sigma_1}} \int \mathrm{d} \mathbf{r} n^{\sigma_1}(\mathbf{r}) n_{\text{xc}, \lambda}^{\sigma_1 \sigma_2}(\mathbf{r}, u).
\end{equation}
The spin $\sigma_1 \sigma_2$ component of the XC energy $E_{\text{xc}}^{\sigma_1 \sigma_2}$ is evaluated by integrating the $\sigma_1 \sigma_2$ system- and spherically-averaged XC hole Eq.~\eqref{eq:spin_nxc} with respect to the inter-electronic distance $u$ and multiplying by the number of spin $\sigma_1$ electrons as
\begin{equation}\label{eq:spin_exc}
    E_{\text{xc}}^{\sigma_1 \sigma_2}\left[n^{\sigma_1}, n^{\sigma_2}\right] = \frac{N_{\sigma_1}}{2} \int_0^1 \mathrm{~d} \lambda \int_0^{\infty} \mathrm{~d} u \frac{4 \pi u^2\langle n_{\text{xc},\lambda}^{\sigma_1 \sigma_2}\rangle(u)}{u}.
\end{equation}
The total system- and spherically-averaged XC hole for an open-shell system must be constructed from the sum of the four individual spin terms $\alpha\alpha$, $\alpha\beta$, $\beta\alpha$ and $\beta\beta$, each evaluated individually according to Eq.~\eqref{eq:spin_nxc}, giving the total expression for the system- and spherically-averaged XC hole as
\begin{equation}\label{eq:nxc_total}
    \begin{small}
    \begin{aligned}
        \langle n_{xc,\lambda}^{}\rangle (u)&= \frac{1}{N_\alpha} \int  \mathrm{d}\mathbf{r} n^\alpha(\mathbf{r}) n_{xc,\lambda}^{\alpha\alpha}(\mathbf{r},u) + \frac{1}{N_\beta} \int  \mathrm{d}\mathbf{r} n^\beta(\mathbf{r}) n_{xc,\lambda}^{\beta\beta}(\mathbf{r},u) \\
        &+ \frac{1}{N_\alpha} \int  \mathrm{d}\mathbf{r} n^\alpha(\mathbf{r}) n_{xc,\lambda}^{\alpha\beta}(\mathbf{r},u) + \frac{1}{N_\beta} \int  \mathrm{d}\mathbf{r} n^\beta(\mathbf{r}) n_{xc,\lambda}^{\beta\alpha}(\mathbf{r},u).
    \end{aligned}
    \end{small}
\end{equation}
Similarly, the total XC energy for an open-shell system evaluated in this way must be constructed from each of the four components $E_{\text{xc}}^{\alpha \alpha}, E_{\text{xc}}^{\alpha \beta}, E_{\text{xc}}^{\beta \alpha}$, and $E_{\text{xc}}^{\beta \beta}$ evaluated according to Eq.~\eqref{eq:spin_exc}, to give the total XC energy as
\begin{equation}\label{eq:exc_total}
    \begin{aligned}
        E_{\text{xc}}^{}\left[n_\alpha, n_\beta\right] &= \frac{N_\alpha}{2} \int_0^1 \mathrm{~d} \lambda \int_0^{\infty} \mathrm{~d} u \frac{4 \pi u^2\langle n_{\text{xc}, \lambda}^{\alpha \alpha}\rangle (u)}{u} \\
        &+\frac{N_\beta}{2} \int_0^1 \mathrm{~d} \lambda \int_0^{\infty} \mathrm{~d} u \frac{4 \pi u^2\langle n_{\text{xc}, \lambda}^{\beta \beta}\rangle (u)}{u} \\
        &+ \frac{N_\alpha}{2} \int_0^1 \mathrm{~d} \lambda \int_0^{\infty} \mathrm{~d} u \frac{4 \pi u^2\langle n_{\text{xc}, \lambda}^{\alpha \beta}\rangle (u)}{u} \\
        &+\frac{N_\beta}{2} \int_0^1 \mathrm{~d} \lambda \int_0^{\infty} \mathrm{~d} u \frac{4 \pi u^2\langle n_{\text{xc}, \lambda}^{\beta \alpha}\rangle (u)}{u}.
    \end{aligned}
\end{equation}
Through spin-scaling relations, exchange hole for a spin-unpolarized density $n_x\left[n\right]$ can be generalized to apply to any spin-polarized density $n_x\left[n^{\sigma_{1}},n^{\sigma_{2}}\right]$ as
\begin{equation}\label{eq:nx_spin_scal}
    n_x \left[n^{\sigma_{1}}, n^{\sigma_{2}}\right](\mathbf{r}, \mathbf{r}+\mathbf{u}) = \sum_\sigma \frac{n^\sigma(\mathbf{r})}{n(\mathbf{r})} n_x\left[2 n^\sigma\right](\mathbf{r}, \mathbf{r}+\mathbf{u}),
\end{equation}
hence it is only necessary to model the exchange hole as a functional of the total density. Furthermore, the exact system- and spherically-averaged exchange and correlation holes satisfy the following sum rules respectively,
\begin{equation}
    \begin{aligned}
        \int^\infty_0\txtd u\, 4\pi u^2 \langle n_\xcx \rangle (u)&=-1,  \\
        \int^\infty_0\txtd u\, 4\pi u^2 \langle\bar{n}_\xcc \rangle (u)&=0. \label{eq:sumrule_XC}
    \end{aligned}
\end{equation}

The CCSD coupling-constant averaged reference XC energy can be obtained directly from one and two particle density matrices: 
\begin{equation} \label{eq:EcCCSD}
E_{\text{xc}}^{\sigma_1 \sigma_2} = \frac{1}{2} \int^1_0 d\lambda\int \mathrm{d}\mathbf{r}_{2}  \int \mathrm{d}\mathbf{r}_{1} \frac{n^{\sigma_1 \sigma_2}_{2,\lambda}\left(\mathbf{r}_{1},\mathbf{r}_{2}\right) - n^{\sigma_{1}}\left(\mathbf{r}_{1}\right) n^{\sigma_{2}}\left(\mathbf{r}_{2}\right)}{\vert\mathbf{r}_{1}-\mathbf{r}_{2}\vert}. 
\end{equation}
This can be used to test the convergence of the XC energies calculated from the system- and spherically-averaged XC holes with respect to $u$, see Eq. ~\ref{eq:exc_total}.

When $\lambda=1$, the correlation energy of a CCSD calculation can be given as:
\begin{equation}
    \epsilon_{c} ^{\sigma_1 \sigma_2} = \frac{1}{2} \int \mathrm{d}\mathbf{r}_{2}  \int \mathrm{d}\mathbf{r}_{1} \frac{n^{\sigma_1 \sigma_2}_{2,\lambda=1}\left(\mathbf{r}_{1},\mathbf{r}_{2}\right) - n^{\sigma_{1}}\left(\mathbf{r}_{1}\right) n^{\sigma_{2}}\left(\mathbf{r}_{2}\right)}{\vert\mathbf{r}_{1}-\mathbf{r}_{2}\vert} - E_x^{HF}. 
\end{equation}
We note that the correlation energy defined above is slightly different from the DFT one in that the orbitals used in $E_x^{HF}$ are not KS orbitals which yield the physical electron density. We therefore use $\epsilon_c$ to differentiate from the $\lambda$-averaged $E_c$ of DFT. The $\epsilon_c$ is only used within Table \ref{tab:cbs/bse} for the purpose of testing the basis set convergence for CCSD.

\section{Computational details}\label{sec:computation}
The computational details of the present work are similar to those outlined in Ref.~\citenum{hou2024capturing}; we used the Lieb optimization method implemented in the \textsc{Quest} \cite{QUEST} code, using a CCSD reference wavefunction and the Gaussian basis expansion of Wu and Yang to optimize the potential.\cite{Wu2003} In this study however we use the spin-unrestricted formalism for all calculations to extend the analysis to open-shell electronic states. We employ Dunning basis sets for all calculations, specifically the d-aug-cc-pVQZ basis set to represent the orbitals and the aug-cc-pVQZ basis set for the potential, using the uncontracted spherical Gaussian form of these basis sets in all cases.~\cite{Dunning1989,Woon1993,Woon1995}
The spherically-averaged XC hole $n_\text{xc}^\lambda$ was constructed by angular integration using an order-41 Lebedev quadrature grid at each reference point~\cite{Lebedev1976,Lebedev1992}. The PBE and LDA XC model holes were calculated with a u interval of 0.01 bohr.


\section{Results and discussion}\label{sec:results}

\subsection{Basis Set Extrapolation}\label{sec:results/bse}
\begin{table*}
\centering
\caption{Basis set convergence analysis for d-aug-cc-pVXZ using the power law extrapolation method~\cite{truhlar1998basis}, with X=D-6 for He and N and X=D-5 for Li, Be, and Ne. The CCSD correlation energy is denoted as $\epsilon^\text{CCSD}_c$, while $\epsilon^\text{(T)}_c$(1) represents the perturbative triple excitation (T) contribution to the correlation energy, the total correlation energy being given by $\epsilon_c$. We use $\epsilon_c$ to differentiate from the $\lambda$-averaged $E_c$ of DFT. The reference HF energies $E_\text{HF}$ and CAS MCHF correlation energies are taken from Ref.~\citenum{davidson1991ground}.}
\label{tab:cbs/bse}
\begin{tabular*}{0.9\linewidth}{@{\extracolsep{\fill}}lllllll@{}}
\toprule
Atom                & Basis Set        & $E_\text{HF}$ / $E_\text{h}$           & $\epsilon^\text{CCSD}_c$ / $E_\text{h}$      & $\epsilon^\text{(T)}_c$ / $E_\text{h}$        & $\epsilon_c$ / $E_\text{h}$          & $E_\text{total}$ / $E_\text{h}$        \\ \midrule
\multirow{7}{*}{He} & d-aug-cc-pVDZ & -2.85573239   & -0.03534795 &             & -0.03534795 & -2.89108034   \\
                    & d-aug-cc-pVTZ & -2.86118442   & -0.03995492 &             & -0.03995492 & -2.90113933   \\
                    & d-aug-cc-pVQZ & -2.86152239   & -0.04118447 &             & -0.04118447 & -2.90270685   \\
                    & d-aug-cc-pV5Z & -2.86162722   & -0.04160561 &             & -0.04160561 & -2.90323283   \\
                    & d-aug-cc-pV6Z & -2.86167323   & -0.04179494 &             & -0.04179494 & -2.90346816   \\
                    & CBS/BSE       & -2.86211189   & -0.04243506 &             & -0.04243506 & -2.90454696   \\
                    & \textbf{Ref~\citenum{davidson1991ground}}         & \textbf{-2.86168}      &             &             & \textbf{-0.04204}    & \textbf{-2.90372}      \\ \midrule 
\multirow{6}{*}{Li} & d-aug-cc-pVDZ & -7.36462240   & -0.02953624 & -0.00001493 & -0.02955117 & -7.39417356   \\
                    & d-aug-cc-pVTZ & -7.43270628   & -0.03754452 & -0.00002408 & -0.03756860 & -7.47027487   \\
                    & d-aug-cc-pVQZ & -7.43271947   & -0.03979260 & -0.00002649 & -0.03981909 & -7.47253856   \\
                    & d-aug-cc-pV5Z & -7.43274670   & -0.04066065 & -0.00003020 & -0.04069085 & -7.47343755   \\
                    & CBS/BSE       & -7.44185253   & -0.04216398 & -0.00003061 & -0.04219459 & -7.48404714   \\
                    & \textbf{Ref~\citenum{davidson1991ground}}         & \textbf{-7.432730}     &             &             & \textbf{-0.045330}   & \textbf{-7.478057}     \\ \midrule
\multirow{6}{*}{Be} & d-aug-cc-pVDZ & -14.57238670  & -0.07898150 & -0.00036407 & -0.07934557 & -14.65173227  \\
                    & d-aug-cc-pVTZ & -14.57287634  & -0.08570531 & -0.00043926 & -0.08614457 & -14.65902091  \\
                    & d-aug-cc-pVQZ & -14.57296957  & -0.08779773 & -0.00055479 & -0.08835252 & -14.66132209  \\
                    & d-aug-cc-pV5Z & -14.57301266  & -0.08905158 & -0.00058869 & -0.08964027 & -14.66265294  \\
                    & CBS/BSE       & -14.57311099  & -0.09005238 & -0.00058697 & -0.09063935 & -14.66375034  \\
                    & \textbf{Ref~\citenum{davidson1991ground}}         & \textbf{-14.57302}     &             &             & \textbf{-0.09434}    & \textbf{-14.66736}     \\ \midrule
\multirow{7}{*}{N}  & d-aug-cc-pVDZ & -54.39339243  & -0.13885654 & -0.00167501 & -0.14053155 & -54.53392399  \\
                    & d-aug-cc-pVTZ & -54.40122030  & -0.16133350 & -0.00284838 & -0.16418188 & -54.56540218  \\
                    & d-aug-cc-pVQZ & -54.40383973  & -0.16866385 & -0.00316775 & -0.17183160 & -54.57567133  \\
                    & d-aug-cc-pV5Z & -54.40447066  & -0.17247573 & -0.00329317 & -0.17576890 & -54.58023957  \\
                    & d-aug-cc-pV6Z & -54.40453785  & -0.17427399 & -0.00334435 & -0.17761834 & -54.58215619  \\
                    & CBS/BSE       & -54.40581459  & -0.17634285 & -0.00350029 & -0.17984314 & -54.58565773  \\
                    & \textbf{Ref~\citenum{davidson1991ground}}         & \textbf{-54.40093}     &             &             & \textbf{-0.18831}    & \textbf{-54.58924}     \\ \midrule
\multirow{6}{*}{Ne} & d-aug-cc-pVDZ & -128.49653420 & -0.27867301 & -0.00412515 & -0.28279816 & -128.77933230 \\
                    & d-aug-cc-pVTZ & -128.53330100 & -0.33258150 & -0.00579453 & -0.33837603 & -128.87167700 \\
                    & d-aug-cc-pVQZ & -128.54376810 & -0.35601024 & -0.00631658 & -0.36232682 & -128.90609490 \\
                    & d-aug-cc-pV5Z & -128.54678850 & -0.36700350 & -0.00657293 & -0.37357643 & -128.92036490 \\
                    & CBS/BSE       & -128.55415230 & -0.37433478 & -0.00684791 & -0.38118269 & -128.93533500 \\
                    & \textbf{Ref~\citenum{davidson1991ground}}         & \textbf{-128.54709}    &             &             & \textbf{-0.39047}    & -\textbf{128.93756}    \\
                    \bottomrule
\end{tabular*}
\end{table*}

To confirm the basis sets used are sufficiently complete, we examine the basis-set convergence of the Hartree-Fock and CCSD correlation energies, using the d-aug-cc-pVXZ series of Dunning basis sets to determine the complete basis set (CBS) limit using the extrapolation power laws,\cite{truhlar1998basis}
\begin{equation}\label{eq:bse}
    \begin{aligned}
    E^{\mathrm{HF, X}} &= E_{\infty}^{\mathrm{HF}} + A^{\mathrm{HF}} \textit{X}^{-\alpha}, \\
    \epsilon^{\mathrm{cor, X}} &= \epsilon_{\infty}^{\mathrm{cor}} + A^{\mathrm{cor}} \textit{X}^{-\beta},
    \end{aligned}
\end{equation}
where \textit{X} represents the cardinal number of the basis set. The values $\alpha = 3.4$ and $\beta_{\mathrm{CCSD}} = \beta_{\mathrm{CCSD}(\mathrm{T})} = 2.4$ are the exponents determined for basis set extrapolation of the HF and CCSD(T) energies respectively in Ref.~\citenum{truhlar1998basis}. The constants $A^{\mathrm{HF}}$ and $A^{\mathrm{cor}}$ are eliminated by solving Eqs.~\eqref{eq:bse} simultaneously for basis sets with two different cardinal numbers. 

The Hartree-Fock, CCSD and perturbative triple excitation (T) correlation energies for this series of basis sets are given in Table~\ref{tab:cbs/bse}, alongside the CBS extrapolated values for each of the five atoms. Note these calculations are conducted for the physical systems with $\lambda = 1$, and thus don't involve the Lieb optimization. The d-aug-cc-pVQZ basis set recovers between 99.85\% (Li) and 99.98\% (Be) of the CBS value for the total energy, and between 94.37\% (Li) and 97.48\% (Be) of the CBS value for the correlation energy. In addition, the (T) contribution to the total CCSD(T) correlation energy is below 2\% at the CBS limit for all atoms considered here. From these results, we consider the CCSD wavefunction in the d-aug-cc-pVQZ basis set to provide adequate accuracy to be used in the Lieb optimization to compute reference XC holes for these atoms. Furthermore, our CCSD(T) CBS energies closely match the Complete Active Space Multi-Configurational Hartree Fock (CAS MCHF) reference values acquired from Ref.~\citenum{davidson1991ground}; the small difference in correlation energy would lead to the CCSD correlation hole being slightly shallower than the exact correlation hole.


\subsection{Convergence Analysis}\label{subsec:convergence}
An essential measure for evaluating the accuracy and reliability of computational methods used to calculate the XC hole is its convergence with $u$, both in recovering the XC energy via Eq.~\eqref{eq:exc_total} and in satisfying the sum rule constraints of Eq.~\eqref{eq:sumrule_XC}. To quantify the convergence of the XC energy, we consider the difference between the value of the XC energy given by Eq.~\eqref{eq:exc_total}, integrated up to some value of $u$, and that evaluated from the CCSD electron density, denoted as $|\int_{0}^{u} \text{d}u' 2\pi {u'} \langle n_{\text{x/c}}^{\text{Method}} (u') \rangle  - E_{\text{x/c}}^{\text{Method}}/N|$. For the XC hole shape convergence, we consider the difference between the integral of the exchange/correlation holes up to some value of $u$ and the theoretical sum rule given by Eq.~\eqref{eq:sumrule_XC}; the differences between these are given by $\left|\int_{0}^{u} \text{d}u' 4\pi {u'}^2 \langle n_{\text{x}}^{\text{Method}} (u') \rangle  + 1\right|$ and $\left|\int_{0}^{u} \text{d}u' 4\pi {u'}^2 \langle n_{\text{c}}^{\text{Method}} (u') \rangle \right|$ for the exchange and correlation holes respectively. 

The reference for the convergence behaviour for both the XC hole shape and XC hole energy is provided by the Lieb+CCSD $\lambda$-integrated system- and spherically-averaged holes. Throughout the convergence testing process, we maintain a fixed interval of 0.01 Bohr for $u$, while exploring the range of 0 to 15 Bohr for Li atom and 0 to 10 Bohr for the other atoms with the Lieb+CCSD method and the range from 0 to 100 Bohr for testing the model exchange/correlation holes. This allows us to gain a complete picture of the convergence properties of these quantities whilst avoiding unnecessary computational expense in constructing the Lieb+CCSD reference values over large values of $u$ beyond those at which the integrals have converged.



\subsubsection{Exchange hole convergence}\label{subsubsec:xholeconv}
\begin{figure*}[!htp]\centering
\includegraphics[width=\textwidth]{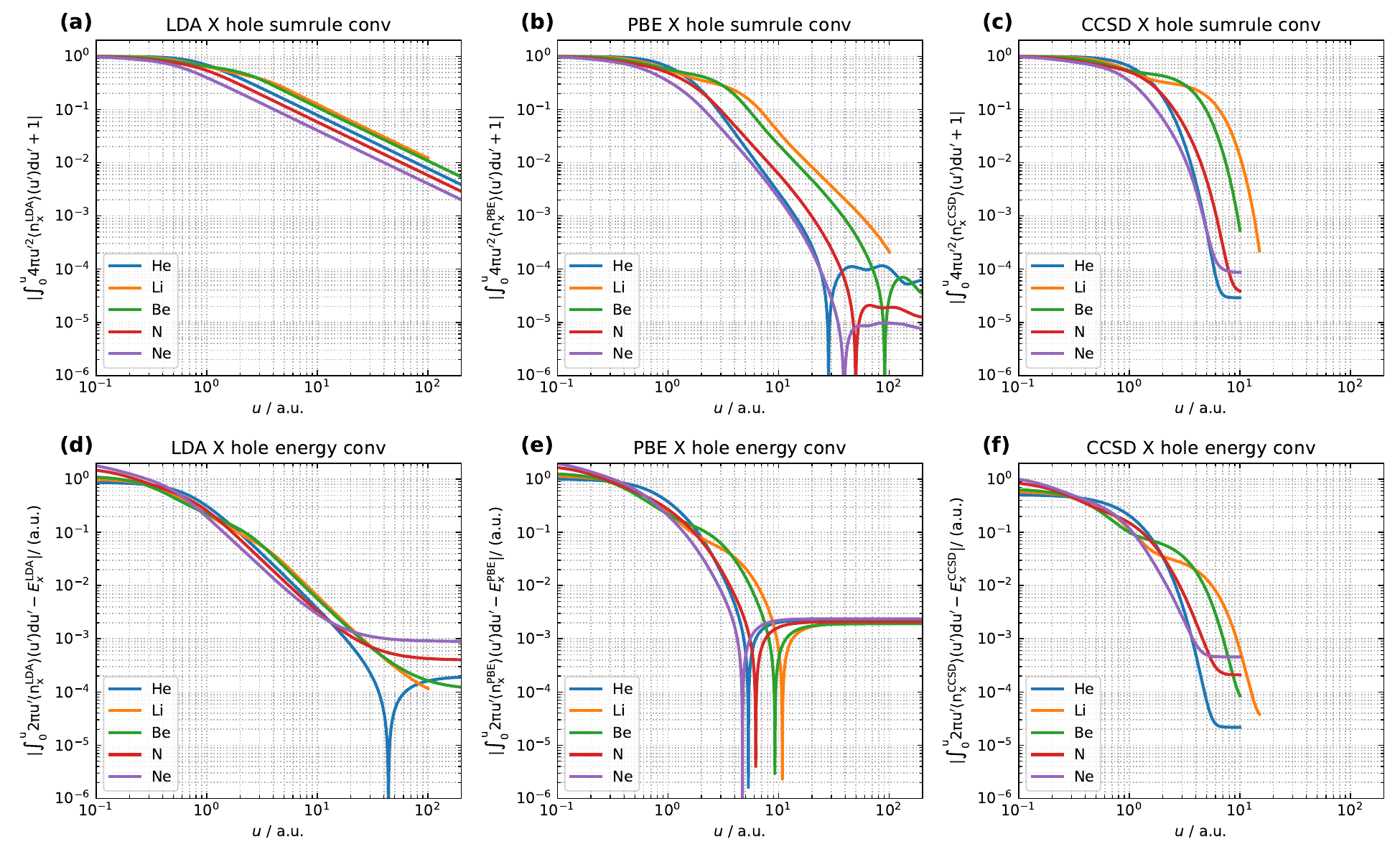}
\caption{The sum rule error and $E_\text{x}$ convergence for the system- and spherically-averaged LDA, PBE and Lieb+CCSD exchange holes for the He, Li, Be, N, and Ne atoms, in the d-aug-cc-pVQZ orbital basis set and the aug-cc-pVQZ potential basis set. }
\label{fig:Xloglog}
\end{figure*}

The convergence measures for the energy and sum rule are presented for the LDA and PBE exchange hole models \cite{ernzerhof1998generalized} alongisde the Lieb+CCSD exchange holes for all five atoms in Fig.~\ref{fig:Xloglog}. 

It can be seen in Fig.~\ref{fig:Xloglog}(c) that, for the Lieb+CCSD exchange holes, the sum rule integrals exhibit relatively rapid convergence with increasing $u$, with the sum rule integral error converging to $10^{-3}$ or below for all atoms at the maximum values of $u$ examined here. The convergence is more rapid for the He, N and Ne atoms, with the sum rule error falling below 10\% by $u = 3$ Bohr, than for the Be and Li atoms, which reach the same level of convergence at $u = 5$ and $u = 7$ Bohr respectively. The sum rule convergence error reaches 1\% at $u = 5$ Bohr for He, N and Ne and at $u = 8$ and $u = 10$ Bohr for Be and Li respectively, whilst the values become $u = 6$, $9$ and $12$ Bohr respectively for the error to fall below 0.1\%. 

Regarding the convergence of the exchange energy, shown in Fig.~\ref{fig:Xloglog}(f) for the Lieb+CCSD exchange holes, the exchange energy per electron is recovered to within $0.1 E_h$ for all atoms in the range $u < 2$ Bohr. Convergence to within $0.01 E_h$ is reached within $u = 2$ Bohr for He, N and Ne whilst for Be and Li this convergence is achieved at $u = 5$ Bohr and $u = 6$ Bohr respectively. To reach convergence of $0.001 E_h$, the range of $u$ must be extended to $5$, $7$ and $9$ Bohr for these systems respectively. 

The convergence of the LDA model exchange holes, shown in Fig.\ref{fig:Xloglog}(a), is much slower than that of the Lieb+CCSD exchange holes. The sum rule integral error declines approximately exponentially, shown by the near-linear decay on the log-log scale of Fig.\ref{fig:Xloglog}(a) and with convergence reached of between 5\%-0.4\% for these atoms at $u=100$ Bohr. The exchange energy per electron converges to within $0.1 E_h$ for all atoms in the range $u < 3$ Bohr and to within $0.01 E_h$ in the range $u < 8$ Bohr. To reach convergence of $0.001 E_h$ requires integration up to $30$ Bohr for He, N and Ne and up to $50$ Bohr for Be and Li. The convergence of the exchange energy per electron with respect to $u$ exhibits a near-linear trend on the log-log axes of Fig.~\ref{fig:Xloglog}(d) in the range $1 < u < 10$ Bohr, tailing off to become approximately constant for Be and Li in the range $20 < u < 100$ Bohr. The data for the He atom exhibits a feature not observed in the other atoms; the convergence becomes much tighter and hits a singularity between $40-50$ Bohr, with the error becoming greater again at larger distances. The sharpness of this minimum is accentuated by the use of logarithmic scales, with a similar effect seen in other plots discussed below. 

For the PBE exchange hole model, the sum rule integral error is shown in Fig.~\ref{fig:Xloglog}(b), in which it can be seen that the behaviour is somewhere between that of the Lieb+CCSD and LDA exchange holes. 
Convergence to within 0.1\% is reached for the He, N and Ne atoms by $u = 20$ Bohr, while the values for Be and Li are $40$ and $60$ Bohr respectively. Notably, for large values of $u$ representing long-range behavior, He, Be, N, and Ne exhibit a significant drop in sum rule integral error convergence, followed by an increase in the error at increasing $u$, although convergence remains below $10^{-4}$. The convergence of the exchange energy per electron evaluated with the PBE model exchange hole, shown in Fig.~\ref{fig:Xloglog}(e), resembles that of the Lieb+CCSD reference in the range $1 < u < 4$ Bohr, with sharp drops in the error being observed for all atoms in the range $4 < u < 12$, with the error then increasing for all atoms with further increases in $u$ to reach an asymptotic level of $2 mE_h$. 


The difference in behaviour of the LDA and PBE exchange hole convergence may be rationalized by considering their shape functions, given by Eq.~(18) and Eq.~(24) of Ref.~\citenum{ernzerhof1998generalized} respectively. The LDA exchange hole shape function is constructed such that the resulting model exchange hole satisfies the sum rule over all space, however as $u\to\infty$ the LDA exchange hole integrand varies as $\propto -1/u^2$. The resulting long-range behaviour of the LDA model exchange hole model leads to the slow convergence of the sum rule error. In comparison, the PBE exchange hole shape function varies as $\propto -e^{-f(s)u^2}/u^2$ as $u\to\infty$, where $f(s)$ is some function of the reduced density gradient $s$, resulting in increasing attenuation of the exchange hole as $u$ increases and more rapid convergence of the sum rule error. The exchange energy converges much more rapidly than the sum rule with $u$ in both cases, which may be understood by comparing Eq.~\eqref{eq:spin_exc} with Eq.~\eqref{eq:sumrule_XC}; in the former, the exchange hole is multiplied by $u$ in the integrand whilst in the latter it is multiplied by $u^2$. 

Overall, the convergence of the sum rule error is more rapid for the Lieb+CCSD exchange holes than for either the LDA or PBE model exchange holes, whilst the convergence of the exchange energies is somewhat similar for the Lieb+CCSD and PBE exchange holes but slower for the LDA exchange hole. 

\subsubsection{Correlation hole convergence}\label{subsubsec:choleconv}
\begin{figure*}[!htp]\centering
\includegraphics[width=\textwidth]{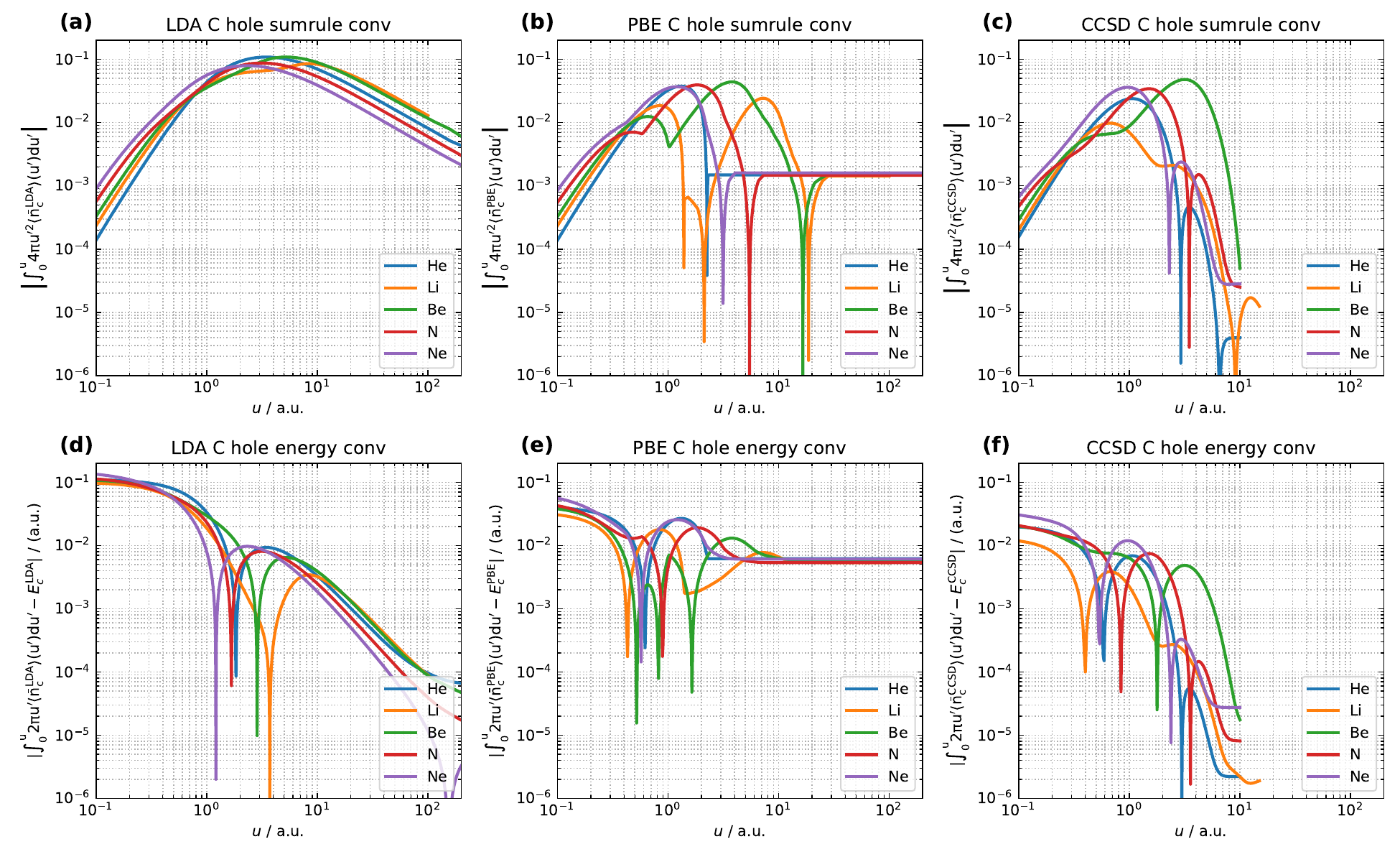}
\caption{The sum rule and $E_\text{c}$ convergence for the system- and spherically-averaged LDA, PBE and $\lambda$-averaged Lieb+CCSD correlation holes for the He, Li, Be, N, and Ne atoms, in the d-aug-cc-pVQZ orbital basis set and the aug-cc-pVQZ potential basis set.}
\label{fig:Cloglog}
\end{figure*}

In an analogous manner to that for the exchange hole, the sum rule and correlation energy convergence for the LDA and PBE model correlation holes are plotted alongside those for the Lieb+CCSD $\lambda$-averaged correlation holes for the five atoms in Fig.~\ref{fig:Cloglog}. 

For the reference Lieb+CCSD correlation hole integrals, shown in Fig.~\ref{fig:Cloglog}(c), it can be seen that the value of the integral at $u=0$ in each case is equal to the on-top correlation hole magnitude, which for these atoms decreases with increasing atomic number: He > Li > Be > N > Ne. With increasing $u$, the magnitude of the correlation hole integral then increases for all atoms but reaches a maximum value and begins to decrease again at the first node of the correlation hole. The position at which this occurs is different for each atom, occurring at $u \sim 0.6$ Bohr for Li, $u \sim 1-2$ Bohr for He, N and Ne and $u \sim 3$ Bohr for Be. The value of the integral at this point depends on the depth of the correlation hole up to the first node; this is shallow in Li thus the maximum value of the correlation hole integral is $\sim 0.01$, whereas the greater depth of the correlation hole in the other atoms results in a maximum of $\sim 0.02-0.05$. Beyond this point, the correlation hole becomes positive causing the integral to decrease in magnitude with increasing $u$. Subsequent maxima in the integral may be attributed to the dominance of positive correlation hole at long range, amplified by the $u^2$ prefactor in the integrand, however these are at least an order of magnitude smaller than the initial maxima. There is a general convergence of the correlation hole integral to below $10^{-4}$ within 10 Bohr for all atoms, with the slowest convergence exhibited by Be for which the value of the integral remains above $10^{-3}$ in the range $u < 8$ Bohr.  

The convergence of the correlation energy evaluated with the Lieb+CCSD correlation holes are shown in Fig.~\ref{fig:Cloglog}(f). In all cases, these exhibit a complicated pattern of convergence as $u$ increases. The value at $u=0$ is simply $\vert E_{c}^{CCSD}\vert$ and there is a general trend for this to converge at large $u$, reaching $0.2mE_{h}$ per electron or below at $u = 10$ Bohr. In-between these two values however, the deviation from $E_{c}^{CCSD}$ for each atom exhibits at least one sharp minimum and one maximum; the positions of the maxima coincide with those in the integral of the correlation hole whilst the sharp minima indicate the recovery of the correlation energy $ E_{c}^{CCSD}$ of Eq. \ref{eq:EcCCSD} from the correlation hole calculations, see Eq. \ref{eq:exc_total}. Since the correlation hole goes above zero as $u$ increases, there can be multiple sharp minima for one system, e.g., He, N, and Ne, as shown in Fig.~\ref{fig:Cloglog}(f).

The convergence of the LDA correlation hole integral towards the sum rule is shown in Fig.~\ref{fig:Cloglog}(a). The values at $u=0$ are again those of the on-top correlation holes for these atoms, which may be more accurate than those of the Lieb+CCSD reference since the inter-electronic cusp is more fully represented in the density functional models.~\cite{hou2024capturing} The correlation hole integrals in the region $u=0.1-0.4$ Bohr follows the trend of those for the Lieb+CCSD reference, reaching a maximum value in the region $u=2-7$ Bohr for all atoms. However, the value of this maximum is greater at around $0.1$ and the subsequent decrease in the integral with increasing $u$ is much slower than for the Lieb+CCSD reference, being near-linear on the log-log scale of Fig.~\ref{fig:Cloglog}(a). This behavior indicates that the LDA model correlation holes are more localized and short-ranged than the reference Lieb+CCSD correlation holes.

Regarding correlation energy convergence, shown in Fig.~\ref{fig:Cloglog}(d), the LDA correlation hole model yields significantly higher values at short-range $u \sim 0.1$ Bohr, compared to Lieb+CCSD. Sharp minima in the correlation energy integration error are present around $u = 2$, $4$, $3$, $2$ and $1$ Bohr for He, Li, Be, N and Ne respectively - indicative of a coincidental agreement of the correlation energy integral, in the region where the correlation hole is negative, and the total correlation energy. As $u$ further increases, the error in the integrated correlation energy returns to the previous trend, decreasing near-linearly on the log-log scale in the range $u \sim 10-100$ Bohr, reaching an error in the order of $0.1mE_{h}$ per electron at $u=100$ Bohr.

The convergence of the PBE model correlation hole to the sum rule is shown in Fig.~\ref{fig:Cloglog}(b). The initial values closely resemble those of the LDA correlation hole, with the two models giving a very similar representation of the inter-electronic cusp region. The increases in the integrals then somewhat resembles those of the Lieb+CCSD reference, although with apparent discontinuities in the derivatives of the curve for N and Be appearing at around $u=0.7$ and $1.0$ Bohr respectively. The difference in the PBE correlation hole integral and that of the Lieb+CCSD reference for Li is significant beyond around $1.0$ Bohr, with two sharp decreases between $u = 1.0-1.1$ Bohr followed by a large increase to above 0.02 at around $u \sim 8$ Bohr. These features are not present in the Lieb+CCSD reference for Li. A notable difference between the PBE correlation hole integral and that of both LDA and Lieb+CCSD is that the PBE correlation hole integral reaches a value of around $1.5\times10^{-3}$ for all atoms with increasing $u$, at which it remains fixed as $u$ is increased up to $100$ Bohr. The discontinuities and the residual at large $u$ are likely due to the cutoff step function used in the PBE correlation hole model to enforce the sum rule \cite{Perdew1996}.


A similar picture is seen for the convergence of the PBE correlation energy in Fig.~\ref{fig:Cloglog}(e). Although the correlation energies are between those of LDA and of Lieb+CCSD and the initial trend at small $u$ then follows that seen for the Lieb+CCSD correlation energy convergence, there are larger number of apparent discontinuities in the derivative of these curves and, by $u = 10$ Bohr, the errors in the correlation energy for all atoms converge to around $6mE_{h}$ beyond which they remain fixed. This is because that the PBE correlation hole model is constructed to always satisfy the sum rule, but it doesn't reproduce the correlation energy functional exactly \cite{Perdew1996}, which can be seen in Table \ref{tab:energies} by comparing $E_c$ from the PBE correlation energy functional and those from the PBE correlation hole model.




\begin{table*}[!htp]
\caption{A comparison of $E_\text{xc}$ values calculated with the LDA and PBE functionals that are evaluated on the CCSD density non self-consistently (DFT@CCSD), and those with the LDA and PBE XC hole models also evaluated non self-consistently on the CCSD density, against values obtained by integrating across the adiabatic connection constructed by Lieb optimization with a CCSD reference, using the d-aug-cc-pVQZ orbital basis set and the aug-cc-pVQZ potential basis set. All energies are given in $E_{h}$. 
}
\label{tab:energies}
\begin{tabular*}{0.9\linewidth}{@{\extracolsep{\fill}}l *{4}{S[table-format=-1.4]} *{2}{S[table-format=-1.4]} *{4}{S[table-format=-1.4]}@{}}
\toprule
Atom & \multicolumn{4}{c}{Non self-consistent DFT@CCSD} & \multicolumn{4}{c}{XC hole model@CCSD} & \multicolumn{2}{c}{Lieb+CCSD} \\ 
\cmidrule(lr){2-5} \cmidrule(lr){6-9} \cmidrule(lr){10-11}
& {LDA $E_\text{x}$} & {LDA $E_\text{c}$} & {PBE $E_\text{x}$} & {PBE $E_\text{c}$} & {LDA $E_\text{x}$} & {LDA $E_\text{c}$} & {PBE $E_\text{x}$} & {PBE $E_\text{c}$} & {$E_\text{x}$} & {$E_\text{c}$} \\
\midrule
He & -0.8829 & -0.1123 & -1.0127 & -0.0418 & -0.8832 & -0.1124 & -1.0149 & -0.0480 & -1.0241 & -0.0417 \\
Li & -1.5374 & -0.1508 & -1.7568 & -0.0513 & -1.5372 & -0.1510 & -1.7598 & -0.0594 & -1.7797 & -0.0401 \\
Be & -2.3203 & -0.2251 & -2.6441 & -0.0861 & -2.3201 & -0.2252 & -2.6479 & -0.0977 & -2.6730 & -0.0894 \\
N & -5.8963 & -0.4268 & -6.5478 & -0.1798 & -5.8949 & -0.4269 & -6.5552 & -0.1985 & -6.5971 & -0.1704 \\
Ne & -11.0158 & -0.7416 & -12.0486 & -0.3493 & -11.0113 & -0.7416 & -12.0606 & -0.3796 & -12.0783 & -0.3594 \\
\bottomrule
\end{tabular*}
\end{table*}

\subsection{Coupling-constant averaged XC holes and hole models}\label{subsec:DFT_XCholes}
In Section~\ref{subsec:convergence} we analysed the exchange and correlation holes of LDA, PBE and Lieb+CCSD by examining their convergence properties - key quantitative measures of the accuracy of model exchange and correlation holes. In this Section, we will extend this analysis to consider their real-space pictures, providing specific insight into how the model exchange and correlation holes may be improved. 

\subsubsection{The exchange hole}
\begin{figure*}[!htp]\centering
\includegraphics[width=\textwidth]{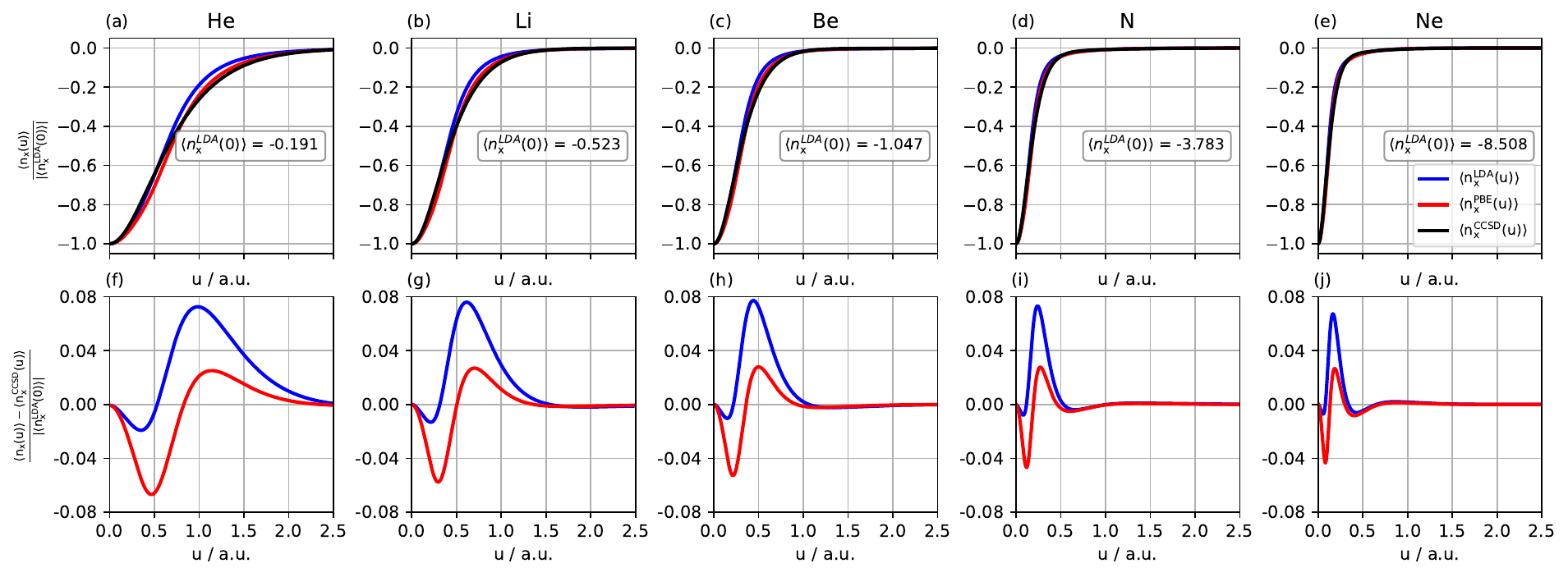}
\caption{The normalized system- and spherically-averaged exchange holes and the normalized difference in the LDA and PBE exchange holes from the Lieb+CCSD reference for the He, Li, Be, N, and Ne atoms, evaluated with the d-aug-cc-pVQZ orbital basis set and the aug-cc-pVQZ potential basis set.}
\label{fig:ccsd_hx}
\end{figure*}

A normalized system- and spherically-averaged exchange hole may be defined as $\langle n_\text{x}\rangle (u)/|\langle n^{\text{LDA}}_\text{x}\rangle(0)|$, where $n^{\text{LDA}}_\text{x}(0)$ is the on-top LDA exchange hole which is known to be exact for systems represented by a single determinant.~\cite{ernzerhof1998generalized} This normalized hole is a useful quantity with which we can easily assess the similarities and differences between models and is plotted for each atom in the upper panels (a)-(e) of Fig.~\ref{fig:ccsd_hx}. It can be seen that all three exchange holes are equal at $u=0$ and that, on this scale, appear to have similar shapes. 

The difference of the normalized LDA and PBE exchange holes with respect to the Lieb+CCSD reference are plotted in the lower panels (f)-(j) of Fig.~\ref{fig:ccsd_hx}, revealing that the normalized LDA exchange hole model exhibits a maximum difference of 0.07-0.08 and, as the number of electrons increases, the value of $u$ at which the greatest positive value of this difference occurs becomes smaller, decreasing from from $1.0$ Bohr in He to $0.2$ Bohr in Ne, indicating the exchange holes become more localized. On the other hand, the first minimum of difference varies from -0.02 to -0.005, approaching zero as the number of electrons increases. Overall, Fig.~\ref{fig:ccsd_hx} (f)-(j) show that the LDA exchange holes are shallower than the Lieb-CCSD reference exchange holes.


In contrast, it can be seen in Fig.~\ref{fig:ccsd_hx}(f)-(j) that the normalized PBE exchange hole model exhibits significantly more negative differences from the reference compared to the normalized LDA exchange hole model, indicating that the PBE exchange hole is deeper than the LDA exchange hole. The negative difference with the largest magnitude is around $-0.06$ for He, decreasing in magnitude as the number of electrons increases to around $-0.04$ for Ne. Meanwhile the largest positive difference remains constant at around $0.02$ for each of the five atoms, with the value of $u$ at which it occurs decreasing from around $1.1$ Bohr to $0.2$ Bohr from He to Ne. Overall, the PBE exchange holes still have considerable deviations from the Lieb+CCSD exchange holes, but these are much more evenly balanced between positive and negative differences, whilst overall being smaller than those of LDA. Consequently, the PBE exchange hole model outperforms the LDA exchange hole model in terms of $E_\text{x}$, as demonstrated in Table~\ref{tab:energies}.

\subsubsection{The correlation hole}

\begin{figure*}[!htp]\centering
\includegraphics[width=\textwidth]{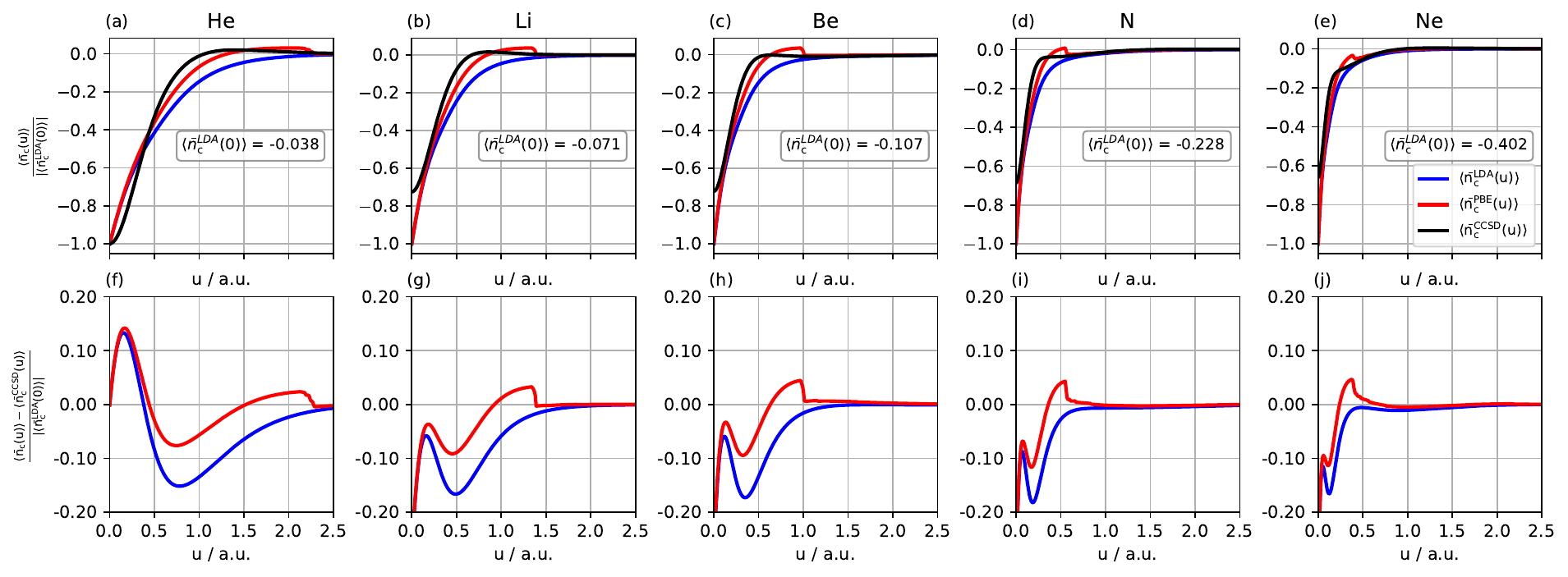}
\caption{The normalized system- and spherically-averaged correlation holes and the normalized difference in the LDA and PBE correlation holes from the $\lambda$-averaged Lieb+CCSD reference for the He, Li, Be, N, and Ne atoms, evaluated with the d-aug-cc-pVQZ orbital basis set and the aug-cc-pVQZ potential basis set.}
\label{fig:ccsd_hc}
\end{figure*}

Similar to the exchange hole, the normalized correlation hole provides a convenient measure for comparing different correlation hole models. It is calculated as $\langle n_\text{c}\rangle (u)/|\langle n^\text{LDA}_\text{c}\rangle(0)|$, where the LDA correlation hole model provides an accurate description of the on-top correlation hole value $n^\text{LDA}_\text{c}(0)$ \cite{burke1998semilocal}. The normalized LDA, PBE and Lieb+CCSD correlation holes are shown in Fig.~\ref{fig:ccsd_hc}(a)-(e), in which we observe that only for He does the on-top correlation hole of Lieb+CCSD match the value obtained from the DFT correlation hole models. For the other atoms, the normalized Lieb+CCSD on-top correlation hole exhibits deviations of approximately 0.3 from the DFT models. As the number of electrons increases, these deviations become more pronounced due to the increasing cusp effect. This suggests that utilizing DFT correlation hole models to correct the cusp effect-driven error in wavefunction correlation would hold promise as a strategy for the future development of high-accuracy electronic structure methods.

The differences in the normalized correlation holes with respect to Lieb+CCSD are shown in Fig.~\ref{fig:ccsd_hc}(f)-(j), clearly revealing an initial region corresponding to the characteristic radius (identified by the first peak) of the cusp effect described in Ref.~\citenum{hou2024capturing}. Notably, radius of this cusp-effect appears slightly smaller for the LDA correlation hole model than the PBE correlation hole model. Beyond this cusp-effect region, the Lieb+CCSD correlation hole can be considered to be more reliable than the DFT models for these systems, as shown in Ref.~\citenum{hou2024capturing}. In Fig.~\ref{fig:ccsd_hc}(f)-(j), for values of $u$ beyond the cusp-effect region, the DFT correlation hole models exhibit predominantly negative differences with respect to the Lieb+CCSD reference, most notably for the LDA correlation hole model. The most negative value of this difference is approximately $-0.17$, arising from the LDA model correlation hole at $u=0.2$ in the N atom. The PBE correlation hole model shows negative differences from the reference at similar values of $u$ to the LDA correlation hole model, but with a magnitude in the order of half that for LDA models.

With the exception of the He atom near the cusp, the LDA correlation hole model yields negative differences with respect to Lieb+CCSD across the full range of $u$ for all atoms, indicating that the LDA correlation hole model is too deeper with respect to the reference one. On the other hand, the PBE correlation hole model exhibits positive differences with respect to Lieb+CCSD in all atoms, with the maximum value increasing from $0.02$ to $0.05$ from He to Ne and the value of $u$ at which this occurs reducing from $2.2$ Bohr to $0.4$ Bohr. These observations indicate that the PBE correlation hole model provides a more accurate estimation of $E_\text{c}$ compared to the LDA correlation hole model, as demonstrated in Table~\ref{tab:energies}. Due to the step-function present in the definition of the PBE correlation hole, the normalized PBE correlation hole exhibits a discontinuity at some value of $u$ for all atoms, with this value becoming smaller with increasing atomic number.

\subsubsection{The XC hole}
\begin{figure*}[!htp]\centering
\includegraphics[width=\textwidth]{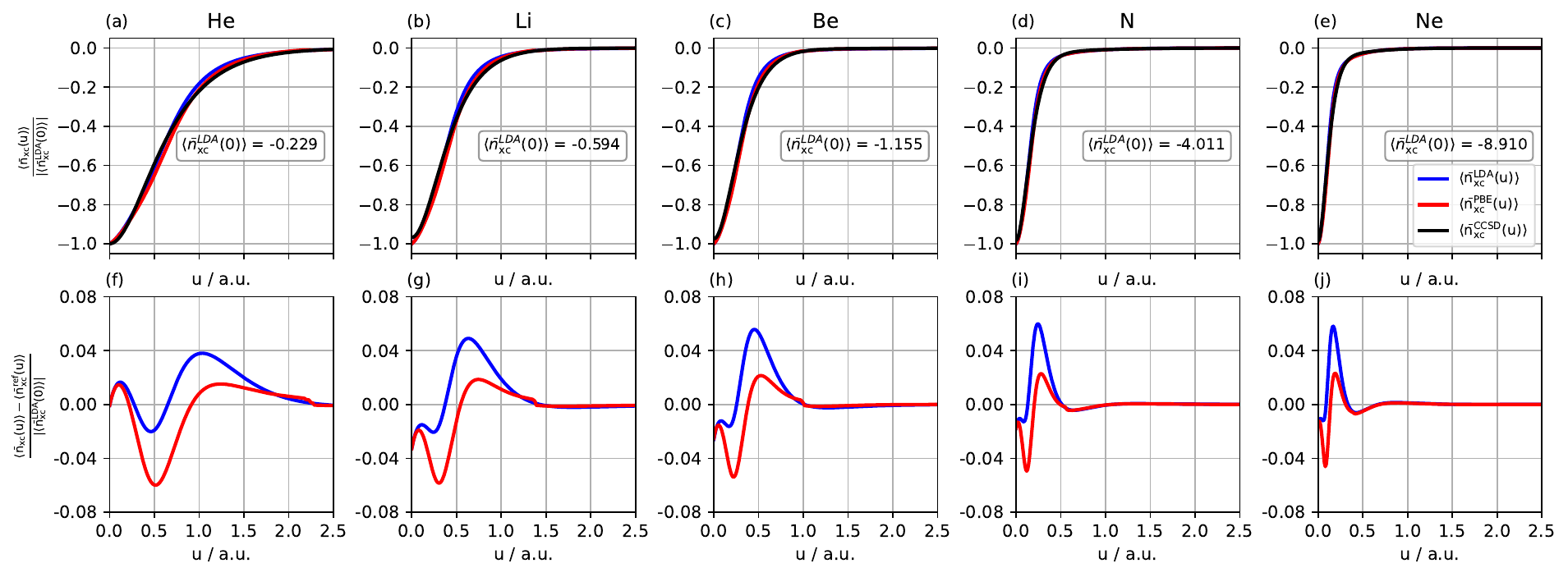}
\caption{The normalized system- and spherically-averaged XC holes and the normalized difference in the LDA and PBE XC holes from the $\lambda$-averaged Lieb+CCSD reference for the He, Li, Be, N, and Ne atoms, evaluated with the d-aug-cc-pVQZ orbital basis and aug-cc-pVQZ potential basis.}
\label{fig:ccsd_hxc}
\end{figure*}

From the previous discussion analysing the exchange and correlation holes individually, we can observe a qualitative trend indicating that the PBE XC hole model has smaller errors in $E_\text{x}$ and $E_\text{c}$ separately compared to the LDA XC hole model. Previous analysis also indicates that the LDA exchange holes are shallower than the Lieb+CCSD reference exchange holes, while the LDA correlation holes are deeper than the corresponding reference ones. Therefore, smaller deviations of the LDA exchange correlation holes from the reference Lieb+CCSD exchange-correlation holes are observed in Fig.~\ref{fig:ccsd_hxc}(f)-(j), illustrated by the smaller peaks of around 0.04 in comparison with those of around 0.08 given in Fig.~\ref{fig:ccsd_hx}(f)-(j). This demonstrates the error cancellation between the LDA exchange and correlation. Similar error cancellations are observed for the PBE exchange and correlation hole models as shown in Figs.~\ref{fig:ccsd_hx}, ~\ref{fig:ccsd_hc}, and ~\ref{fig:ccsd_hxc}, but to a lesser extent since each component individually provides a closer representation of the Lieb+CCSD exchange and correlation holes than those of the LDA models.

\section{Conclusion}\label{sec:conclusion}
In this work, we present the system- and spherically-averaged XC holes for the helium, lithium, beryllium, nitrogen and neon atoms at the CCSD level of accuracy, constructed across AC using the Lieb optimization method and $\lambda$-averaged to provide an accurate benchmark for DFT XC hole models outside of the inter-electronic cusp region. In addition, this approach yields accurate values of $E_\text{x}$ and $E_\text{c}$ against which those evaluated with DFT models may be compared. 

We present a thorough numerical analysis of the Lieb+CCSD exchange and correlation holes in comparison to the LDA and PBE model exchange and correlation holes. We examine the convergence behaviour of their integrals with respect to inter-electronic separation to analyse their adherence to the sum rules for exchange and correlation holes and the recovery of the exchange and correlation energies from the respective holes. Whilst the Lieb+CCSD exchange holes demonstrate rapid convergence with respect to $u$ both in respect of the sum rule and exchange energy, the LDA and PBE exchange hole models show a more gradual convergence with increasing $u$ and do not reach the same levels of convergence as the Lieb+CCSD reference in the same range. Examination of the shape functions for the LDA and PBE model exchange holes allows the improved convergence properties of the PBE model exchange hole to be rationalized. 

We observe a different picture for the convergence behaviour of the correlation holes compared with that for the exchange holes. The convergence of the LDA correlation hole sum rule exhibits a near-linear trend on the log-log scale as $u$ becomes large, deviating significantly from the Lieb+CCSD reference. The PBE correlation hole sum rule is more similar to the Lieb+CCSD reference at low $u$ but features many apparent derivative discontinuities at increasing $u$, tending to a constant value for all atoms at large $u$ attributed to the presence of a step-function in the PBE correlation hole model. It is important to note that the convergence of the correlation energy differs from that of the sum rule, with sharp points of apparently tight convergence as the value of the integral approaches that of the total correlation energy in the region where the correlation hole is negative. 

We concluded our analysis with an examination of the real-space characteristics of the normalized LDA and PBE exchange and correlation holes, in comparison to those of the Lieb+CCSD reference. We observe that both the exchange and correlation holes become more localized as the number of electrons increases. The exchange hole shape remains similar across all atoms, while the first maximum in the correlation hole decreases in magnitude as the number of electrons increases. Furthermore, we found that there is generally greater cancellation of errors between the LDA exchange and correlation hole models than for PBE.

In summary, our analysis provides valuable insights into the characteristics and performance of exchange, correlation and combined XC hole models, underlining the importance that such models are carefully assessed to enhance the accuracy and reliability of DFT calculations in which they are used. In the future, we aim to improve the accuracy of the XC hole benchmark by combining the LDA correlation hole model around the inter-electronic cusp region with the Lieb+CCSD correlation hole beyond this region. This combined approach holds the promise of creating a more comprehensive representation of the XC hole, opening the door to the development of improved XC hole models and ultimately advancing the the accuracy and reliability of DFT calculations.

\section{Acknowledgments}
This work was supported by National Science Foundation (NSF) under Grant No. DMR-2042618. A.M.T. and T.J.P.I. are grateful for support from the European Research Council under H2020/ERC Consolidator Grant “topDFT” (Grant No. 772259). This work was supported by the Norwegian Research Council through CoE Hylleraas Centre for Quantum Molecular Sciences Grant No. 262695.

\bibliography{aipsamp}
\end{document}